\documentclass[         %
aps,                    
prd,                    
superscriptaddress,    
nofootinbib,            
preprintnumbers,        %
floatfix]               
{revtex4}               
\usepackage{graphicx,longtable,amsmath}

\numberwithin{equation}{section}

\begin{document}

\title{Neutrino Mixing and Oscillations in Astrophysical Environments}

\author{A.B. Balantekin}
\affiliation{Physics Department, University of Wisconsin, Madison WI 53706 USA}

\begin{abstract}
A brief review of the current status of neutrino mixing and oscillations in astrophysical environments, with particular emphasis on the Sun and core-collapse supernovae, is given. Implications of the existence of sterile states which mix with the active neutrinos are discussed. 
\end{abstract}

\maketitle


\section{Introduction}

Neutrinos play a key role in the formation and the evolution of the cosmos. Consequently all the properties of neutrinos (masses, mixing angles, 
CP-violation nature) could significantly impact description of astrophysical environments. The fact that neutrino mass and weak-interaction eigenstates do not coincide (but are related with a unitary transformation) is a well-established experimental fact.  
From the measurements of the invisible decay width of Z, we know that only three (active) flavors couple to the weak interactions,
Solar, atmospheric and reactor experiments were able to obtain two of the mixing angles and the two differences between squares of the neutrino masses. The remaining mixing angle, $\theta_{13}$, was recently measured with very high precision both in the disappearance experiments with reactor neutrinos 
\cite{An:2012eh,Ahn:2012nd,An:2012bu,An:2013zwz,Abe:2012tg} and in the appearance experiments with accelerator neutrinos 
\cite{Abe:2013hdq,Adamson:2013ue}. The values of any CP-violating phases in the mixing matrix are not yet measured. The hierarchy of the neutrino masses, whether neutrinos mix with particles which do not interact weakly themselves or violate the total lepton number are still open questions. 
A recent review of the current status of the neutrino oscillations in given in Ref. \cite{Balantekin:2013tqa}. to which the reader is referred to for further details. 

In these conference proceedings we briefly review of the current status of neutrino mixing and oscillations in astrophysical environments, with particular emphasis on the Sun and core-collapse supernovae. We also discuss the interpretation of various anomalous results in neutrino experiments as active-sterile mixing and the implications of this intrepretation. 

\section{Solar Neutrinos}
Before its operations were terminated the Sudbury Neutrino Observatory (SNO) was able to perform a low-threshold analysis down to an effective electron kinetic energy of 3.5 MeV \cite{Aharmim:2009gd}. The result 
was reported as the electron-neutrino survival probability expanded around the maximum of the neutrino energy spectrum of $^8$B solar neutrinos, 
$E_{\nu} = 10$ MeV. It turns out that the solar electron-neutrino survival probability is doubly degenerate in $\delta m^2_{21}$ \cite{Balantekin:2011ft}: one solution with $\delta m^2 \sim 7 \times 10^{-5}$ eV$^2$, historically called LMA (large mixing angle) solution, and another solution 
with $\delta m^2  \sim 10^{-7}$ eV$^2$, historically called LOW (low mass-squared difference) solution. 
SNO low-energy threshold analysis also provides a fit to the day-night asymmetry, which is consistent with zero when the statistical and systematic errors are taken into account.  Including this asymmetry in the analysis does not remove the degeneracy mentioned above; SNO data alone permits both the LOW and LMA values of $\delta m^2_{21}$. Including recent SuperKamiokande results does not alter this conclusion \cite{Abe:2010hy}.  
Of course the KamLAND experiment clearly picks the LMA solution \cite{Abe:2008aa}. 
However, to establish the equivalence of the neutrino parameters for neutrinos and antineutrinos, an independent verification of the LMA value  using {\em neutrinos}, instead of reactor {\em antineutrinos} is necessary. One way to do so is to measure the day-night asymmetry for lower energy neutrinos, where the asymmetry is expected to be very large for the LOW solution. Recently the Borexino experiment reported a measurement of the day-night asymmetry for the $E_{\nu} = 0.86$ MeV $^7$Be line neutrinos \cite{Bellini:2011yj}. 
The reported value of the asymmetry, A = 0.001 $\pm$ 0.012 (stat) $\pm$ 0.007 (syst), completely rules out the LOW region.

With the increasingly high precision of the solar neutrino experiments, it is now possible to use solar neutrinos as observational probes of the Sun and tools to aid its modeling. A careful analysis of the physics input into modeling the Sun complements such an endeavor. A critical discussion of the nuclear physics input into solar modeling with the aim of determining the best values and uncertainties on the relevant S-factors was published in 1998 \cite{Adelberger:1998qm}. This analysis was recently updated to provide a set of standard S-factors and uncertainties that reflect the progress made since the previous effort \cite{Adelberger:2010qa}. There has also been considerable recent progress in modeling the Sun, especially its metallicity (abundances of elements heavier than hydrogen and helium). An improved analysis of the solar abundances gives the ratio of solar metallicity to the hydrogen abundance to be $(Z/X)_{\odot} = 0.0178$   \cite{Asplund:2009fu}, as compared to the previously accepted value of $(Z/X)_{\odot} = 0.0229$ \cite{Grevesse:1998bj}. Using these ratios and the recent reevaluation of the nuclear reaction rates in the Sun \cite{Adelberger:2010qa}, new solar model calculations were carried out \cite{Serenelli:2011py}. These recent calculations suggest that the difference between old and new metallicities is hard to distinguish using already measured pp chain neutrino fluxes, however they significantly change the values of the CNO neutrino fluxes.  

When the neutrinos travel through the matter, the standard free particle relation 
\begin{equation}
\label{1}
E^2 = \mathbf{p}^2 + m^2
\end{equation}
is modified to read
\begin{equation}
\label{2}
(E-V)^2= (\mathbf{p} - \mathbf{A})^2 + m^2, 
\end{equation}
where the scalar ($V$) and vector ($\mathbf{A}$) corrections arising from the neutrino interactions with the background particles are integrated and averaged over this background. Since many background particles would contribute to such an averaging, these corrections will depend on the macroscopic variables associated with the background.  For the scalar correction these are the matter density and electron fraction. The vector ($\mathbf{A}$) corrections can be proportional to the vector variables such as the large-scale currents present or the total spin in case of polarized backgrounds. 
In the limit of static, charge-neutral and unpolarized backgrounds, the vector corrections vanish and the scalar correction due to the standard weak interactions is given by the Wolfenstein potentials \cite{Wolfenstein:1977ue}
\begin{equation}
\label{3}
V_e = \sqrt{2} G_F N_e (x) - \frac{1}{\sqrt{2}}G_F N_n (x)
\end{equation}
and 
\begin{equation}
\label{4}
V_{\mu} = V_{\tau}=  - \frac{1}{\sqrt{2}}G_F N_n (x),
\end{equation}
where $N_e$ and $N_n$ are the electron and neutron densities of the medium, respectively. (Since the medium is taken to be charge neutral, the proton density is equal to the electron density, $N_p = N_e$ and is not explicitly written), The potentials in Eq. \ref{3} and \ref{4} are provided by the coherent forward scattering of neutrinos off the  electrons, neutrons, and protons in dense matter. The term  $- G_F N_n/\sqrt{2}$ is  due to Z-exchange. Since it is the same for all neutrino flavors, it does not contribute to phase differences unless there are sterile neutrino states. These potentials give rise to the MSW effect \cite{Wolfenstein:1977ue,Mikheev:1986gs} for neutrino oscillations in matter. Note that matter effects induce an effective CP-violation since the matter in the Earth and in the stars is not CP-symmetric. 

The assumption that the background is static, charge-neutral and unpolarized may not always hold in astrophysical settings. For example, in many cases strong magnetic fields may be present in the stars and other compact objects, impacting neutrinos \cite{Semikoz:1987py}. 
Indeed neutrino propagation and oscillations are quite different in a polarized medium as compared to the unpolarized case 
\cite{Nunokawa:1997dp}. Similarly large-scale matter currents in the stars can give rise to very interesting effects \cite{Cumming:1996gv}. 

Many times calculations exploring the neutrino propagation in matter treats the background static, but even in the absence of polarization and large scale currents matter density may fluctuate.  
The role of density fluctuations in the Sun were first examined two decades  
ago \cite{Loreti:1994ry}. This is an area which nicely illustrates that in the 
interdisciplinary area of neutrino physics and astrophysics, an outlook from 
a nuclear physics perspective could be uniquely beneficial: introduction of 
density fluctuations in the Sun and supernovae borrows tools developed in the 
compound nucleus theory. Using the data from the solar neutrino and KamLAND experiments stringent limits were later given  
on solar density fluctuations \cite{Balantekin:2003qm}. These  
neutrino data constrain solar density fluctuations to be less than 
$\beta = 0.05$ at the 70 \% confidence level, where $\beta$ is the fractional 
fluctuation around the value given by the Standard Solar Model, with the best fit to the combined solar neutrino and KamLAND data 
given by $\beta = 0$. Even though one does not expect a fluctuation as large 
as five percent, such analyses illustrate a proof of principle, namely that 
it is now possible to probe {\em solar} physics with the neutrino data. 

\section{Supernova Neutrinos}

Some epochs of the Early Universe and core-collapse supernovae can be viewed as dynamical systems dominated by neutrinos. 
A recent brief review on the role of neutrinos in these environments is given in Ref. \cite{Balantekin:2013gqa}. 
99\% of the energy emitted following the collapse of the progenitor star is emitted in the form of neutrinos and antineutrinos of all flavors. In addition core-collapse supernovae may host several nucleosynthesis sites: r-process in the neutrino-driven wind from the surface of the proto-neutron star and the $\nu$-process in the outer shells. Because of the sheer number of neutrinos present $(\sim 10^{58})$, several novel aspects of the neutrino physics come into play. For example, one can no longer ignore the interactions between neutrinos themselves. The resulting collective neutrino oscillations dominate neutrino propagation in the vicinity of the proto-neutron star. Details of such collective effects are recently reviewed in Refs. 
\cite{Duan:2010bg}, \cite{Raffelt:2010zza} and \cite{Duan:2009cd}. Neutrino-neutrino interactions lead to novel collective and emergent effects, such as conserved quantities and interesting features in the neutrino energy spectra (called spectral "swaps" or "splits" 
\cite{Raffelt:2007cb,Duan:2008za,Pehlivan:2011hp,Galais:2011gh}). In fact, there is a duality between the Hamiltonian describing neutrino propagation in matter including neutrino-neutrino interactions and the BCS Hamiltonian of the superconducting many-body systems; the same symmetry present in both cases leads to analogous dynamics \cite{Balantekin:2006tg,Volpe:2013uxl}. 
This symmetry naturally leads to splits in the neutrino energy spectra and it was used to find conserved quantities and constants of motion in the limit where the angles between momenta of the neutrinos are the same \cite{Pehlivan:2011hp}. Collective neutrino oscillations are usually studied assuming that neutrinos always scatter in the forward direction. However neutrinos can also scatter in non-forward directions creating a 
"neutrino halo". A significant number of the outgoing neutrinos could interact with this halo \cite{Cherry:2012zw}. 

A persistent problem in the numerical treatment of collective oscillations is the current-current nature of the weak interactions at low energies, introducing a term 
\begin{equation}
\label{5}
\left( 1 - \frac{\mathbf{p}_i}{p_i} \cdot \frac{\mathbf{p_j}}{p_j} \right) 
\end{equation}
into the two-body interaction between neutrinos labeled $i$ and $j$ in the many-neutrino Hamiltonian. In Eq. (\ref{5}) $\mathbf{p}_i$ and $\mathbf{p}_j$ are the momenta of those neutrinos. Many times the angle between these momenta are taken to be the same as was done in deriving the invariants mentioned above. In the case of supernovae, typically axially symmetric neutrino emission is assumed and the azimuth angle is integrated over. 
However, including the azimuth angle of neutrino propagation as an explicit variable leads to a multi-azimuth-angle instability \cite{Raffelt:2013rqa}. 
Models incorporating this instability can be used to demonstrate that solutions of the equations of collective flavor oscillations do not necessarily preserve the symmetries of initial conditions \cite{Raffelt:2013isa}. Numerical solutions of the non-linear neutrino propagation equations in a supernova, introducing the azimuthal angle as angular variable in addition to the usual zenith angle, indeed confirm that flavor conversions depend  on the initial asymmetry between $\nu_e$ and $\bar{\nu}_e$ spectra \cite{Mirizzi:2013rla}. 

In general neutrino transport in matter of arbitrary density requires solution of quantum kinetic equations \cite{Sigl:1992fn,Strack:2005ux}
Ideally quantum kinetic equations govern the evolution of neutrino flavor at any density and temperature. They reduce to the Boltzmann equation in the high density limit and the Hamiltonian formalism we have been discussing so far in the low density limit. There have been encouraging recent efforts towards writing down and solving quantum kinetic equations \cite{Vlasenko:2013fja,Zhang:2013lka}. 

\section{Sterile neutrinos and other heavy neutral particles}

Some of the anomalies observed in neutrino experiments can be interpreted as mixing of  sterile neutrinos with active ones \cite{Abazajian:2012ys}. It was argued that there exists a discrepancy\footnote{This is not a universally agreed conclusion. Although earlier analyses concurred with this assessment (see e.g. \cite{Huber:2011wv}), recent work suggests that nuclear corrections that give rise to this anomaly are very uncertain  
\cite{Hayes:2013wra}. The ultimate resolution of this issue lies in further experiments as one needs to precisely measure any relative distortion of the 
$\bar{\nu}_e$ spectrum as a function of both energy and baseline \cite{Djurcic:2013oaa}.} in reactor neutrino experiments between observed antineutrino fluxes near the reactor core and the predicted values \cite{Mention:2011rk}
This anomaly can be fitted with additional sterile neutrino states with mass splittings of the order of $\sim 1$ eV$^2$ and oscillation lengths of a few meters (see e.g. Ref. \cite{Kopp:2013vaa}).  An interpretation of the LSND \cite{Aguilar:2001ty} and  MiniBooNE \cite{Aguilar-Arevalo:2013pmq}
experiments in terms of active-sterile oscillations with a $\sim 1$ eV$^2$ mass-difference scale requires a mixing angle of the order of $\sin^2 2\theta \sim 10^{-2}-10^{-3}$, somewhat lower than what is needed to explain the proposed reduction of reactor neutrino flux. In addition a single sterile state cannot explain possible differences between $\nu_{\mu} \rightarrow \nu_e$ and  $\bar{\nu}_{\mu} \rightarrow \bar{\nu}_e$ channels. 
Global analyses of the short-baseline neutrino oscillation data in 3+1 (three active and one additional sterile state), 3+2 (two additional sterile states both with eV scale masses) and 3+1+1 (two additional sterile states- one at eV scale and one much heavier) neutrino mixing schemes are recently reviewed in Ref. \cite{Giunti:2013waa}. 

\begin{figure}[t]
\label{f:1}
\caption{Schematic depiction on photon coupling to the neutrino mass eigenstates. The coupling strength is usually expressed as neutrino dipole moments.}
\includegraphics[height=.2\textheight]{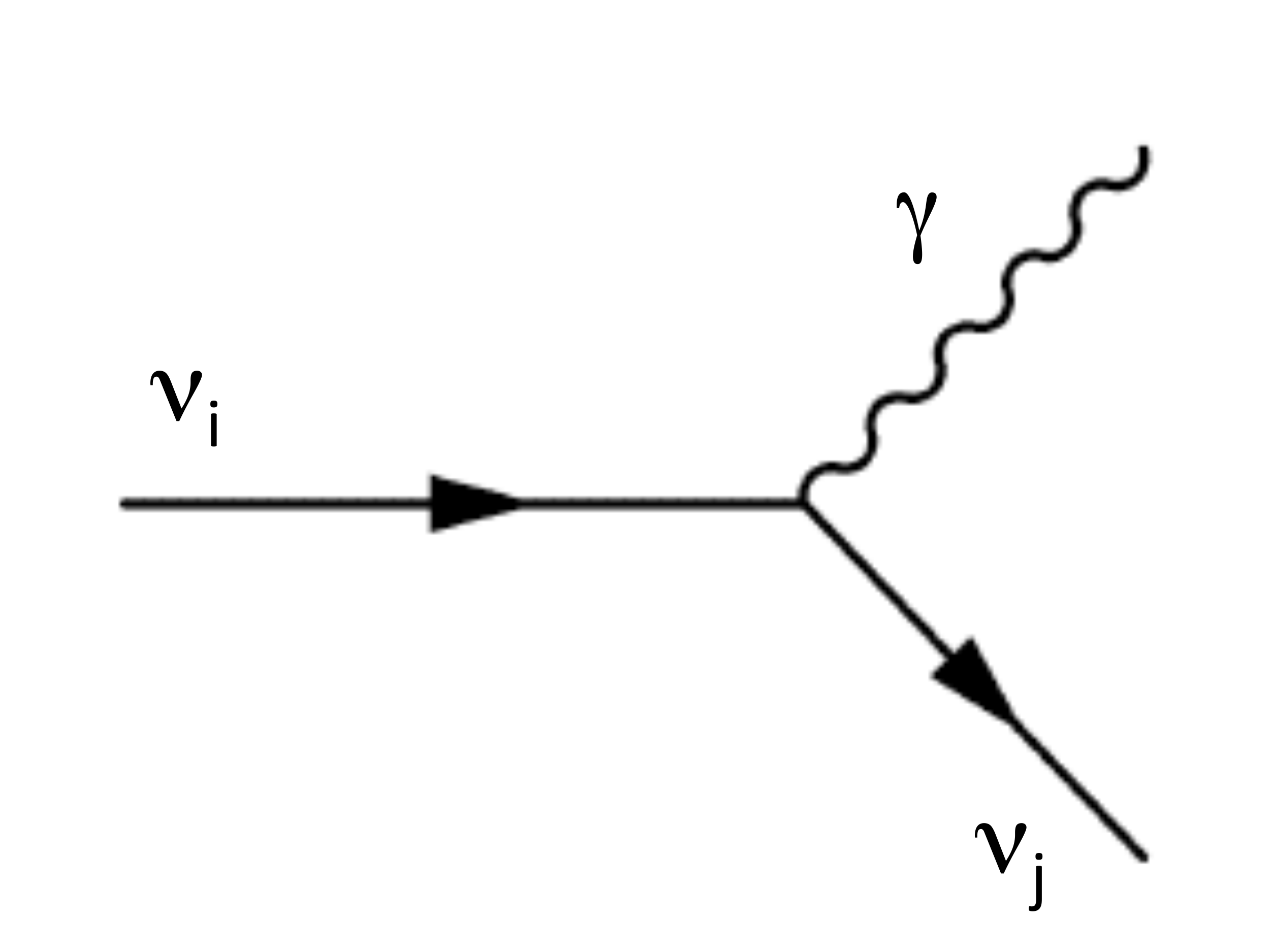}
\end{figure}

The hypothesis of light sterile neutrinos has very interesting consequences in astrophysics\footnote{For example, neutrino oscillations can alter the ratio of electron-neutrino and electron-antineutrino fluxes in a core-collapse supernova, yielding values of the electron fraction dependent on the parameters of the mixing. This in turn would make the yields of the r-process nucleosynthesis depend on the neutrino parameters, should the core-collapse supernovae be the appropriate sites.  It turns out that active-sterile mixing with a mass-difference scale $\sim$ a few eV$^2$ and 
a mixing angle of the order of $\sin^2 2\theta \sim 10^{-2}-10^{-3}$ prevents the electron fraction from getting too high, enabling a robust r-process nucleosynthesis \cite{McLaughlin:1999pd,Caldwell:1999zk}.
(See also Ref.  \cite{Wu:2013gxa} a more recent attempt).}. By definition, sterile states are not produced in weak interactions, but as depicted in Figure \ref{f:1} they can have electromagnetic couplings depending on the physics beyond the Standard Model which would govern their interactions. 
The strength of the coupling is given by the electric and magnetic dipole moments of the neutrinos. 
Such couplings are probed by experiments detecting recoiling electrons in the reactor antineutrino flux.  At low electron recoil energies, the electromagnetic contribution to the cross section could exceed the weak contribution. The current best bound is given by the GEMMA spectrometer at  Kalinin Nuclear Power Plant to be $\mu_{\nu} <  2.9 \times 10^{-11} \mu_B$ at 90\% C.L. \cite{Beda:2013mta}. It was recently pointed out that 
even if the sterile state oscillates and reduces the total flux by the time neutrinos reach the distances where detectors measuring electron recoil are placed, these sterile states can be repopulated after the scattering process \cite{Balantekin:2013sda}. 

Sizable neutrino dipole moments would have interesting astrophysical and cosmological implications. A large enough neutrino magnetic moment implies enhanced plasmon decay rate, $\gamma^* \rightarrow \nu + \bar{\nu}$ in any astrophysical plasma. Since the neutrinos would freely escape the star, for example this process would cool a red giant star faster by delaying the helium ignition. The most recent limit is from the red-giant branch in the globular cluster M5: $\mu_{\nu} < 4.5 \times 10^{-12} \mu_B$ (95$\%$ CL) \cite{Viaux:2013hca}. Both experimental and astrophysical limits quoted here are much larger than the prediction of the minimally expanded Standard Model. Hence a measurement of the neutrino magnetic moment would be a signature of new physics. 

In the standard scenarios for the Big Bang nucleosynthesis, neutrinos are assumed to have decoupled when the temperature drops to 
$T \sim 2 m_e$. After that instant only electron-positron pairs keep interacting with the photons. However, if the dipole moments are non-zero, heavy sterile states could decay into active and other sterile states by the emission of a photon which would not be in thermal equilibrium. Nonthermal photons can induce electromagnetic cascade showers, producing many additional nonthermal photons, albeit less energetic than the original ones (see e.g.,~\cite{Ellis:1990nb,Kawasaki:1994sc}). If the sterile species decay after the $e^+e^-$ annihilation occurs, these nonthermal photons can disintegrate background light elements, potentially altering abundances of nuclei synthesized during the Big Bang nucleosynthesis epoch 
\cite{Cyburt:2002uv,Jedamzik:1999di,Kawasaki:2004qu,Kusakabe:2006hc}. 
Constraints imposed by the observed abundances of all the light elements produced during the Big Bang  
on the decay of the heavy sterile states into non-thermal photons are explored in Ref. \cite{Kusakabe:2013sna}. This mechanism seems to provide a solution of the $^7$Li abundance problem if a sterile state with mass $\sim 3.5$ MeV with a large enough radiative decay width were to exist
\footnote{A similar observation was also made in \cite{Pospelov:2010hj}}.

There is a close coupling between microscopic fundamental physics and macroscopic physics describing the Universe. Neutrinos seem to play a special role in this connection. A through knowledge of the neutrino properties not only will provide us a more complete understanding of the fundamental physics, but it will also enable us to attain a better understanding of the Universe.


This work was supported in part by the U.S. National Science Foundation Grant No.  PHY-1205024, and in part by the University of Wisconsin Research Committee with funds granted by the Wisconsin Alumni Research Foundation. I thank the organizers of OMEG12 for their gracious hospitality. 







\end{document}